\RequirePackage{fix-cm}
\documentclass[smallextended]{svjour3}       
\smartqed  
\usepackage[usenames,dvipsnames]{xcolor}
\usepackage{graphicx}
\usepackage{ogonek}

\usepackage{hyperref}

\usepackage{amssymb}
\usepackage{amsfonts}
\usepackage{bm}
\usepackage{mathrsfs}
\usepackage{mathtools}
\usepackage{amssymb} 

\usepackage{comment}

\usepackage{longtable}

\usepackage[linesnumbered]{algorithm2e}

\usepackage[sans]{dsfont}

\usepackage{fancyvrb}
\usepackage[procnames]{listings}

\mathchardef\mhyphen="2D 

\begin{document}

\title{\textsc{supFunSim}: spatial filtering toolbox for EEG
}


\author{Krzysztof Rykaczewski \and Jan Nikadon \and W{\l}odzis{\l}aw Duch \and Tomasz Piotrowski
}


\institute{Krzysztof Rykaczewski \at
              Center for Modern Interdisciplinary Technologies, Nicolaus Copernicus University, Wile\'nska 4, 87-100 Toru\'n \\
              Faculty of Mathematics and Computer Science, Nicolaus Copernicus University, Chopina 12/18, 87-100 Toru\'n\\
              \email{mozgun@mat.umk.pl}           
          \and
          Jan Nikadon \at
              Center for Modern Interdisciplinary Technologies, Nicolaus Copernicus University, Wile\'nska 4, 87-100 Toru\'n \\
              Faculty of Humanities, Nicolaus Copernicus University, Fosa Staromiejska 1a, 87-100 Toru\'n\\
              \email{nikadon@gmail.com}           
          \and
          W{\l}odzis{\l}aw Duch \at
              Center for Modern Interdisciplinary Technologies, Nicolaus Copernicus University, Wile\'nska 4, 87-100 Toru\'n \\
              Faculty of Physics, Astronomy and Informatics, Nicolaus Copernicus University, Grudzi\k{a}dzka 5/7, 87-100 Toru\'n \\
              \email{wduch@is.umk.pl}           
          \and
          Tomasz Piotrowski \at
	    Center for Modern Interdisciplinary Technologies, Nicolaus Copernicus University, Wile\'nska 4, 87-100 Toru\'n \\
            Faculty of Physics, Astronomy and Informatics, Nicolaus Copernicus University, Grudzi\k{a}dzka 5/7, 87-100 Toru\'n\\
              \email{tpiotrowski@is.umk.pl}           
}

\date{Received: date / Accepted: date}

\maketitle

\begin{abstract}
Recognition and interpretation of brain activity patterns from EEG or MEG signals is one of the most important tasks in cognitive neuroscience, requiring sophisticated methods of signal processing. The \textsc{supFunSim} library is a new \textsc{Matlab} toolbox which generates accurate EEG forward models and implements a collection of spatial filters for EEG source reconstruction, including linearly constrained minimum-variance (LCMV), eigenspace LCMV, nulling (NL), and minimum-variance pseudo-unbiased re\-du\-ced-rank (MV-PURE) filters in various versions. It also enables source-level directed connectivity analysis using partial directed coherence (PDC) and directed transfer function (DTF) measures. The \textsc{supFunSim} library is based on the well-known \textsc{FieldTrip} toolbox for EEG and MEG analysis and is written using object-oriented programming paradigm. The resulting modularity of the toolbox enables its simple extensibility. This paper gives a complete overview of the toolbox from both developer and end-user perspectives, including description of the installation process and some use cases.
\end{abstract}

\section{Introduction}\label{introduction}

Network neuroscience is at present the most promising approach to understand the structure and functions of complex brain networks. Neuroimaging and signal processing methods are rapidly evolving, with the ultimate goal of reaching high time and space resolution, allowing for models of functional connectivity, activation of large-scale networks and their rapid dynamic transitions in multiple time scales. Electroencephalography (EEG) has excellent temporal resolution, is noninvasive and relatively easy to use. Unfortunately, signals observed at the scalp level are difficult to interpret, due to the propagation of signals from their cortical and subcortical sources through multiple layers of the brain with several different volume conduction properties, including the scalp, skull, cerebrospinal fluid (CSF), and brain tissues. Sensors receive  corrupted mixed signals from various active brain structures. Therefore, direct scalp-level EEG analysis cannot reflect the underlying neurodynamics.

Reconstruction of brain's electrical activity from electroencephalographic (EEG) 
or magnetoencephalographic (MEG) recording is frequently based on spatial filters.
\footnote{Spatial filters are also known as beamformers in array signal processing.}
Such task requires solving forward and inverse problems in EEG or MEG and is of 
great interest to EEG/MEG community. Several libraries implementing various
forward and inverse solutions are available, see for example \cite{FieldTrip2011,Brainstorm2011,MNE2014,zumer2014nutmeg,gonzalez2018third} and references therein. 
\textsc{supFunSim} extends their functionality by providing object-oriented implementation of both EEG forward and inverse problems. In the former case, it uses realistic and extendable model of activity of sources and enables usage of accurate EEG forward models. In the latter case, it implements a large collection of spatial filters for EEG source reconstruction, including the linearly constrained minimum-variance (LCMV) 
\cite{Frost1972,VanVeen1997,Sekihara2008}, eigenspace LCMV \cite{Sekihara2008}, nulling (NL)
\cite{Hui2010}, and minimum-variance pseudo-unbiased reduced-rank
(MV-PURE) filters \cite{Piotrowski2008,Piotrowski2009,Piotrowski2019}. It also enables source-level 
directed connectivity analysis using partial directed coherence (PDC) 
\cite{Baccala2001} and directed transfer function (DTF) \cite{Kaminski2001} measures that can be applied to the time series representing reconstructed activity of sources of interest.

The \textsc{supFunSim} toolbox is based on \textsc{FieldTrip}
\cite{FieldTrip2011}, an excellent \textsc{Matlab} toolbox for EEG and
MEG signal analysis. It can be used as an extension (plugin) to \textsc{FieldTrip}. 
Thanks to its object-oriented design, functionality of its modules can be easily extended, e.g., by providing data or functions from other toolboxes.

The source code of the current version of the toolbox is publicly available at 
{\footnotesize\url{https://github.com/nikadon/supFunSim.git}} as an
\lstinline!Org-mode! file, \textsc{Jupyter} notebook, and also as a
plain \textsc{Matlab} source code. Functions are not precompiled, as script libraries have
the advantage of being easily maintainable and extensible.


The paper is organized as follows. First, we briefly introduce the
forward and inverse problems in EEG (similar considerations also apply to MEG). 
Then, we discuss the benefits of the object-oriented approach 
for our toolbox and its extensibility. We close with the use cases which 
appear frequently in practical applications of this toolbox. 
In appendices mathematical details of the implemented spatial filters, 
and the full list of toolbox parameters, are provided.

\section{EEG Measurement Model} \label{forward-solution}

Electromagnetic signals that originate within \emph{enchanted
loom}\footnote{{\em ``The great topmost sheet of the mass, that where
hardly a light had twinkled or moved, becomes now a sparkling field of
rhythmic flashing points with trains of traveling sparks hurrying
hither and thither. The brain is waking and with it the mind is
returning. It is as if the Milky Way entered upon some cosmic dance.
Swiftly the head mass becomes an enchanted loom where millions of
flashing shuttles weave a dissolving pattern, always a meaningful
pattern though never an abiding one; a shifting harmony of
subpatterns''.}\\ \null\hfill{}-- Charles S. Sherrington, Man on his nature. 1942} of
the human brain are propagated through vaious head compartments
\cite{mosher1999eeg}. This \emph{dissolving pattern} of brain
electrical activity can be detected on the surface of scalp using
electroencephalography (EEG).

At a given time instant the EEG data acquisition can be well
approximated by a linear equation of the generic form
\begin{equation} \label{eq:model}
  y = H(\theta)q + H_\mathfrak{i}(\theta_\mathfrak{i})q_\mathfrak{i} + H_\mathfrak{b}(\theta_\mathfrak{b}) q_\mathfrak{b} + n_\mathfrak{m},
\end{equation}
where, for $m$ EEG sensors located at the scalp and $l$
sources of interest modelled as equivalent current dipoles (ECD) located inside
brain's volume,
\begin{itemize}
  \item by $y \in \mathbb{R}^m$ we denote the signal observed
    in the sensor space at a given time instant,
  \item $\theta = (\theta_1, \ldots, \theta_l) \in \Theta \subset
    \mathbb{R}^{l \times 3}$ represents locations of the sources of
    interest, i.e., for the $i\text{-th}$ source the vector
    representing its location is $\theta_i \in \mathbb{R}^3.$ Here,
    $\Theta$ denotes the set of all subsets of locations of source
    signals, 
  \item $H(\theta) \in \mathbb{R}^{m \times l}$ is the sensor array
    response (\textit{lead-field}) matrix of the sources of interest,
  \item $q\in \mathbb{R}^l$ is a vector of electric activity of the
    $l$ sources of interest representing magnitudes of ECDs,
  \item similarly, for the $k$ interfering noise sources,
    $\theta_\mathfrak{i} = \big(\theta^\mathfrak{(i)}_1, \ldots,
    \theta^\mathfrak{(i)}_k\big) \in \mathbb{R}^{k \times 3}$ are the
    interference source locations,
    $H_\mathfrak{i}(\theta_\mathfrak{i}) \in \mathbb{R}^{m \times k}$
    is the corresponding lead-field matrix, $q_\mathfrak{i} \in
    \mathbb{R}^k$ is the corresponding interference activity, 
  \item for the $p$ sources of background activity of the brain
    $\theta_\mathfrak{b} := \big(\theta^\mathfrak{(b)}_1, \ldots,
    \theta^\mathfrak{(b)}_p\big) \in \mathbb{R}^{ p \times 3}$ are the
    background source locations (i.e. sources which are not sources of
    interest and are uncorrelated with them),
    $H_\mathfrak{b}(\theta_\mathfrak{b}) \in \mathbb{R}^{ m \times p
    }$ is the corresponding lead-field matrix, $q_\mathfrak{b} \in
    \mathbb{R}^p$ is the corresponding background activity,
  \item $n_\mathfrak{m} \in \mathbb{R}^m$ is an additive white
    Gaussian noise (AWGN) interpreted as a measurement noise present
    in the sensor space.
\end{itemize}

Equation~\eqref{eq:model} represents a single sample of
EEG data $y$ from subjects' scalp at a given time (\emph{forward
solution}). It enables, through customizable parameters, accurate
modelling of real-world EEG experiments. We shall emphasize at this
point that not all components of the model in \eqref{eq:model} have to
be considered, i.e., one may select only those signal components that
fit the aims of user's simulation settings.

The lead-field matrices establishing signal propagation model are
estimated on the basis of geometry and electrical conductivity of head
compartments together with position of sensors on the scalp. We
consider these properties to be fixed in time during a single EEG data
acquisition session. Therefore, lead-field-matrices are assumed to be
time-invariant in such circumstances (its values do not change during acquisition time in a single session).
We also note that the above EEG forward model (\ref{eq:model}) assumes
that the orientations of the ECD moments are fixed during measurement
period, and only their magnitudes $q$, $q_\mathfrak{i}$, $q_\mathfrak{b}$
vary in time.
We also assume that orientations of the ECD moments are normal and
directed outside with respect to the cortical surface mesh. This is in
accordance with the widely recognized physiological model of EEG
signal origin that considers pyramidal cortical neurons to be the main
contributor to the brain's bioelectrical activity that can be measured
on the human scalp \cite{ElectromagneticBrainMapping962275}.

We assume that $q$ and $q_\mathfrak{i}$ may be correlated (i.e., that sources of interest can interfere with each other), but are uncorrelated with the background sources $q_\mathfrak{b}$ and
the noise $n_\mathfrak{m}.$ We further assume that $q$, $q_\mathfrak{i}$,
$q_\mathfrak{b}$, $n_\mathfrak{m}$ are zero-mean weakly stationary
stochastic processes with the exception that $q$ may contain in
addition a deterministic component simulating evoked (phase-locked)
activity in event-related EEG experiments. In our toolbox the
presence of this component is controlled by the \lstinline!SETUP.ERPs!
variable.

\section{EEG Source Reconstruction} \label{inverse-solutions}

Having solved the EEG forward problem which introduced, in particular,
the lead-field matrices embodying the propagation model of brain's
electromagnetic activity, we are in a position to solve the inverse
problem. Here it amounts to reconstruction of time courses
of activity of sources at predefined locations. That means that we
assume that the locations of the sources of interest $\theta$ are
known. This can be achieved by defining regions of interest using
source localization methods, e.g., minimum-norm
\cite{Pascual-Marqui1999} or spatial filtering-based methods
\cite{Moiseev2011,Piotrowski2018}, or referring to
neuroscience studies that have identified regions of interest (see for example  
Hui et al. \cite{Hui2010}). Then, the goal is to reconstruct the activity $q$ of
sources of interest based on the observed signal $y$ as
\begin{equation}
  \widehat{q} = Wy,
\end{equation}
where $W \in \mathbb{R}^{l \times m}$ is a matrix representing spatial
filter's coefficients. The definitions of the
filters currently implemented in the toolbox are given in Appendix
\emph{Implemented spatial filters}.

\begin{figure}[htbp]
  \centering
  \includegraphics{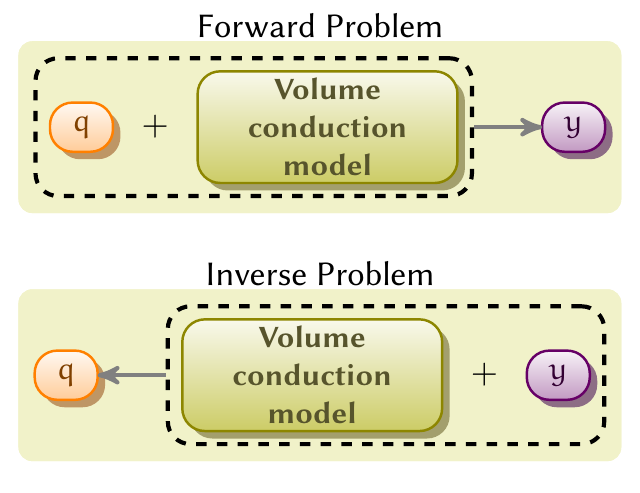}
  \caption{The relationship between forward and inverse problems.}
  \label{fig:inverse}
\end{figure}

\section{Toolbox Signal Processing Outline}
\subsection{Overview}
In order to obtain EEG signal $y$ we need first to generate source activity
signals and propagate them to the position of electrodes, according to the forward model (FM)
given in Equation~\eqref{eq:model}. The source signals are generated
using the method described in \cite{franaszczuk1985application}, which
uses stable multivariate autoregression (MVAR) model with predefined
coefficient matrices. This results in a wide-sense stationary signals
generated with predefined pairwise linear dependencies (correlations).
Such approach has been studied and used in literature, see, e.g.,
\cite{2012_Haufe_TowardsEEGsourceconnectivityanalysis,Baccala2001,2001_Neumaier_Estimationofparametersandeigenmodesofmultivariateautoregressivemodels},
and is especially useful in investigating functional dependencies
between activity of sources using directed connectivity measures such
as partial directed coherence (PDC) \cite{Baccala2001} or directed
transfer function (DTF) \cite{Kus2004}.

Gaussian pseudo-random vectors simulating autoregressive processes are generated using the \lstinline!arsim! function available from
\cite{2001_Schneider_Algorithm808ARfitamatlabpackagefortheestimationofparametersandeigenmodesofmultivariateautoregressivemodels}. 
In this way, we obtain multivariate time series representing activity of
sources of interest $q$ (denoted in the code as
\lstinline!sim_sig_SrcActiv.sigSRC!), sources of interference due to noise
$q_\mathfrak{i}$ (\lstinline!sim_sig_IntNoise.sigSRC!), and sources of
background activity $q_\mathfrak{b}$ (\lstinline!sim_sig_BcgNoise.sigSRC!).


Moreover, as our framework allows to add event-related potentials
(ERP, flag variable \lstinline!SETUP.ERPs! in the code) to the source
signal, each source activity is divided into \emph{pre} and \emph{pst}
parts in relation to the onset of event ($q$ into $q_{pre}$ and $q_{pst}$, etc.) in order to enable
simulation of ERP experiments. In particular, the ERP signal may be
added to $q_{pst}$, but not to $q_{pre}$.

Furthermore, the \emph{pre} and \emph{pst} subsignals are used to implement spatial filters. In particular, noise correlation matrix $N$ may be estimated from $y_{pre}$ signal and signal correlation matrix $R$ may be estimated from $y_{pst}$ signal.

The $y_{pst}$ signal is also used for evaluation of the fidelity of
reconstruction, as a filter $W_f$ produces an estimate of the activity signal
source $\widehat{q}_{pst, f}$ based on $y_{pst}.$ Then, the MVAR model, which is represented by composite model matrix $A00$, is fitted to the
reconstructed source activity using \lstinline!arfit! function,
yielding reconstructed composite MVAR model matrix $A00_f$. This
matrix can then be used to investigate directed connectivity using
partial directed coherence (PDC)~\cite{Baccala2001} and directed transfer function (DTF) \cite{Kaminski2001}.

\begin{figure}[h]
  \centering
  \includegraphics[width=9cm]{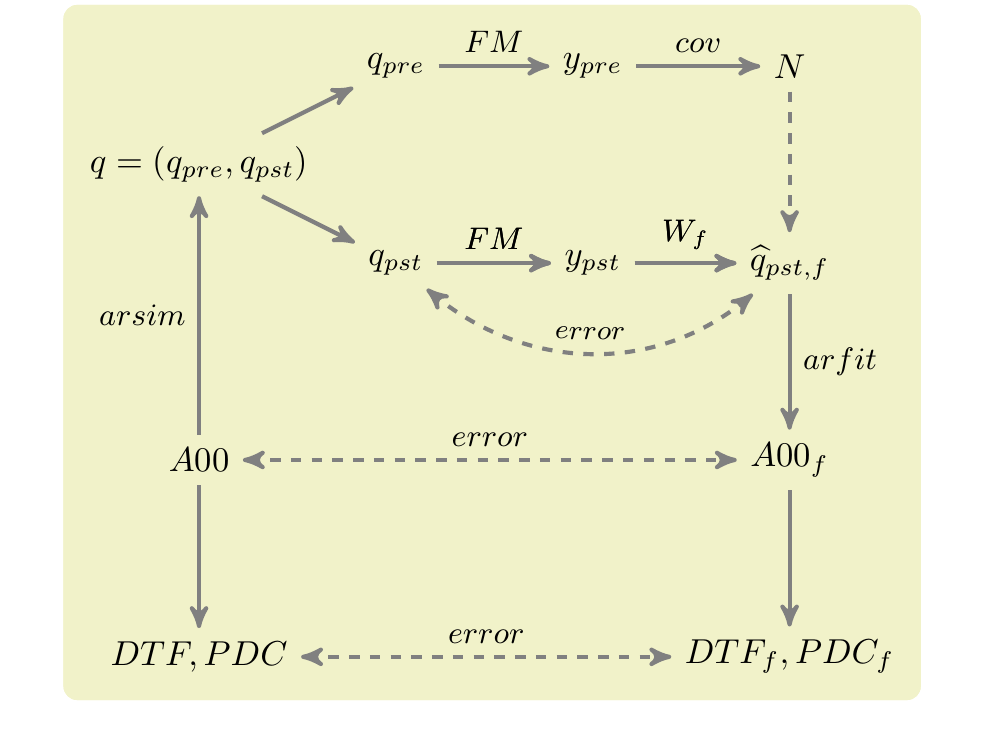}
  \caption{Overview of processing of signals by out toolbox. \lstinline!arsim! and \lstinline!arfit! functions are taken from 
  \cite{2001_Schneider_Algorithm808ARfitamatlabpackagefortheestimationofparametersandeigenmodesofmultivariateautoregressivemodels}, $A00_f$ is the reconstructed composite MVAR model matrix, and $W_f$ is the matrix with spatial filter's coefficients.}
  \label{dt:schemat_all}
\end{figure}

The overview of signals processed by the toolbox and dependencies
between them is depicted in Figure~\ref{dt:schemat_all}.

\subsection{Brain Signals in the Source Space.}
\label{sec:orgf197cd3}

Items \ref{sec:orgdc5cd49} and \ref{sec:orgbbaae1b} below describe 
generation of source signals $q,q_\mathfrak{i}$ and $q_\mathfrak{b}.$
Item \ref{sec:org187472a} concerns definition of volume conduction
model. The computation of lead-field matrices $H$, $H_\mathfrak{i}$ and
$H_\mathfrak{b}$ is discussed in the subsequent Section \emph{Brain
Signals in Sensor Space}.

\begin{enumerate}
  \item \textbf{Positioning sources} \label{sec:orgdc5cd49}

  \smallskip

  File \lstinline!supFunSim/mat/sel_atl.mat! contains 15000 vertex
    FreeSurfer \cite{FreeSurfer1999} brain tessellation together with
    atlases \cite{FreeSurfer1999,FreeSurferCortex2004} that provide
    parcellation of the mesh elements into cortical patches (regions
    of interest, ROIs). This file is provided with the BrainStorm
    toolbox \cite{Brainstorm2011}. First, we randomly select an
    arbitrary number of ROIs by choosing items from Destrieux and
    Desikan-Killiany atlases
    \cite{FreeSurferCortex2004,FreeSurferCortex2006}. In each ROI,
    each vertex is a candidate node for location of the dipole source.
    Then, an arbitrary number of locations can be drawn within each
    ROI, separately for $q$, $q_\mathfrak{i}$ and $q_\mathfrak{b}$.

  Most of the simulation parameters are controlled using \lstinline!SETUP!
  structure. The geometrical arrangement and number of cortical sources
  in each ROIs is controlled using \lstinline!SRCS! field
  (\lstinline!SETUP.SRCS!), which is a three-column \lstinline!<int>!
  array, where:
  \begin{itemize}
    \item rows represent consequent ROIs (thus, the number of
      rows determines the number of ROIs used in the simulations),
    \item the first column represents sources of interest, the second
      column represents sources of interference, and the third column
      represents sources of background activity,
    \item integer values of this array represent number of sources in
      the given ROI for the given signal type.
  \end{itemize}

  For the end-user this provides mechanism not only to control the
  total number of sources of a particular type, but also to choose
  their spatial distribution. Additionally, we provide a mesh
  representing both \emph{thalami} (jointly) as a structure
  containing potential candidates for the non-cortical (deep)
  sources of signal/noise. Variable \lstinline!SETUP! contains also
  the field (\lstinline!SETUP.DEEP!) which defines the number of
  signals arising in the brain center (around thalami) belonging to a
  particular signal type (of interest, interference, or background
  activity). Furthermore, in order to account for the
  mislocalization of sources, together with the original
  lead-fields we also generate \emph{perturbed} lead-fields for the
  source activity reconstruction. These are generated using
  locations that are shifted with respect to the original locations
  in the direction that is rotated in relation to the original (normal
  to cortex surface) dipole orientation. Default shift is random and
  less then $5$ mm in each direction $(x, y, z)$. Default rotation is random, 
  bounded by $\frac{\pi}{32}$ (azimuth and elevation) angle.

  \item \textbf{Sources Timecourse} \label{sec:orgbbaae1b}

  \smallskip
  Following~\cite{2001_Neumaier_Estimationofparametersandeigenmodesofmultivariateautoregressivemodels,2001_Schneider_Algorithm808ARfitamatlabpackagefortheestimationofparametersandeigenmodesofmultivariateautoregressivemodels},
  we use stable 
  autoregressive (MVAR) model to generate time-series. It is assumed
  that such model generates a realistic source
  activity~\cite{korzeniewska2003determination}. We create separate
  models for time-series $q$ and $q_\mathfrak{b}$. The
  $q_\mathfrak{i}$ is obtained as a negative of $q$ with Gaussian
  uncorrelated noise added with the same power as the $q$, i.e.
  $q_\mathfrak{i} = -q + n_\mathfrak{i}$. In this way, we obtain
  correlated time-series $q$ and $q_\mathfrak{i}$.

  The $l\text{-variate}$ autoregressive model of order $p$ for a
  stationary time-series of state vectors ${{q}^{n}} \in \mathbb{R}^l$
  is defined at time instant ${n}$ as 
  \begin{equation}
    {{q}^{(n)}} = \underset{s = 1}{\overset{p}{\mathop \sum }} \, {{A}_{s}}{{q}^{(n - s)}} + {{\varepsilon}_{n}},
    \label{eq:mvarmodel}
  \end{equation}
  where ${{q}^{(n)}}$ is the state vector at time $n$, $p$ is the
  order of the model ($p = 6$ by default), matrices ${{A}_{1}},
  \ldots, {{A}_{p}} \in \mathbb{R}^{l \times l}$ are the coefficient
  matrices of the AR model, and ${{\varepsilon}_{n}}$ is the
  $l$-dimensional additive white Gaussian noise
  \cite{2012_Haufe_TowardsEEGsourceconnectivityanalysis}. For the
  signal of interest $q$, we also give the possibility to include
  deterministic component simulating evoked (phase-locked) activity in
  event-related EEG experiments. The presence of this component is
  controlled by the \lstinline!SETUP.ERPs! variable. Then,
  $q = q^{(n)} + q^{(d)}$, where $q^{(d)}$ is the deterministic ERP component. In the
  toolbox, $q^{(d)}$ is generated using \textsc{Matlab}
  \lstinline!gauswavf! function generating first derivative of Gaussian
  wavelet function, as recommended by~\cite{vrondik2011erp}.

  Similarly, $q_\mathfrak{b}$ is simulated using independent,
  random and stable MVAR model (of order $r = 6$ by default):
  \begin{equation}
    {{q}^{(n)}_\mathfrak{b}} = \underset{s = 1}{\overset{r}{\mathop \sum }} \, {{B}_{s}}{{q}^{(n - s)}_\mathfrak{b}} + {{\varepsilon}_{n}^\mathfrak{b}}.
  \end{equation}
  MVAR model is considered to be \emph{stable} if the absolute values
  of all eigenvalues of all matrices $A_s$ (respectively, $B_s$) are
  less than one. We used procedure adapted
  from~\cite{gomez2008measuring} (namely, we adapted the
  \lstinline!stablemvar! function) to generate stable MVAR model that
  was used for times-series generation. During generation of MVAR
  models for $q$ and $q_\mathfrak{b}$, the coefficient matrix $A_s$
  (respectively, $B_s$) is multiplied by a masking matrix that has
  80\,\% (by default) of its off-diagonal elements equal to zero. All
  the remaining diagonal and off-diagonal masking coefficients are
  equal to one. In code, the composite MVAR model matrix is
  represented by the variable $A00$, see Fig. \ref{fig:a00}.
  Such procedure allows, in particular, to obtain specific profile of
  directed dependencies between activity of sources of interest. This
  approach is taken from~\cite{Baccala2001}. Moreover, it gives
  us the possibility to implement the PDC and DTF directional casual
  dependency measures. Namely, partial directed coherence and directed 
  transfer function are matrices 
  defined using Fourier transform of MVAR model~\eqref{eq:mvarmodel}, i.e.
  \begin{equation}
    A(\lambda) := \mathbb{I}_l - \sum_{s = 1}^{p} A_s \exp(-2\pi i s \lambda),
  \end{equation}
  where $\mathbb{I}_l$ is the identity matrix, $\lambda$ is
  normalized frequency, $|\lambda| \leq 0.5.$ Then, partial directed
  coherence between $i$-th and $j$-th signals is given by
  \cite{Baccala2001}
  \begin{equation}
    \mathrm{PDC}_{ij}(\lambda) := \frac{ A_{ij}(\lambda) }{\sqrt{a^{*}_j(\lambda) a_j(\lambda)}} = \frac{ A_{ij}(\lambda) }{\sqrt{ \sum_{i=1}^{l}\left| A_{ij}(\lambda) \right|^2 }},
  \end{equation}
  where $A_{ij}(\lambda)$ is $ij$ element of matrix $A(\lambda)$, $a_j(\lambda)$ is
  $j$th column of $A(\lambda)$ and $*$ means Hermitian transpose. It takes
  values in the interval $[0, 1]$ and measures the relative strength of
  the interaction of a given source signal $j$ to source signal $i$
  normalized by strength of all of $j$'s connections to other signals
  \cite{blinowska2011practical}.

  Directed Transfer Function (DTF) is defined as
  \begin{equation}
      \mathrm{DTF}_{ji}(\lambda) := \frac{ H_{ji}(\lambda) }{\sqrt{\sum_{i=1}^{l}\left| H_{ji}(\lambda) \right|^2}},
  \end{equation}
  where $H(\lambda) := \big(I - A(\lambda)\big)^{-1}$ is the \textbf{transfer
  matrix}, $i, j= 1, \ldots, l$. It can be interpreted as ratio between
  inflow from channel $i$ to channel $j$ normalized by the sum of inflows to
  channel $j$.

\begin{figure}[htbp]
  \centering
  \includegraphics[width=\columnwidth, height=30mm, keepaspectratio=false]{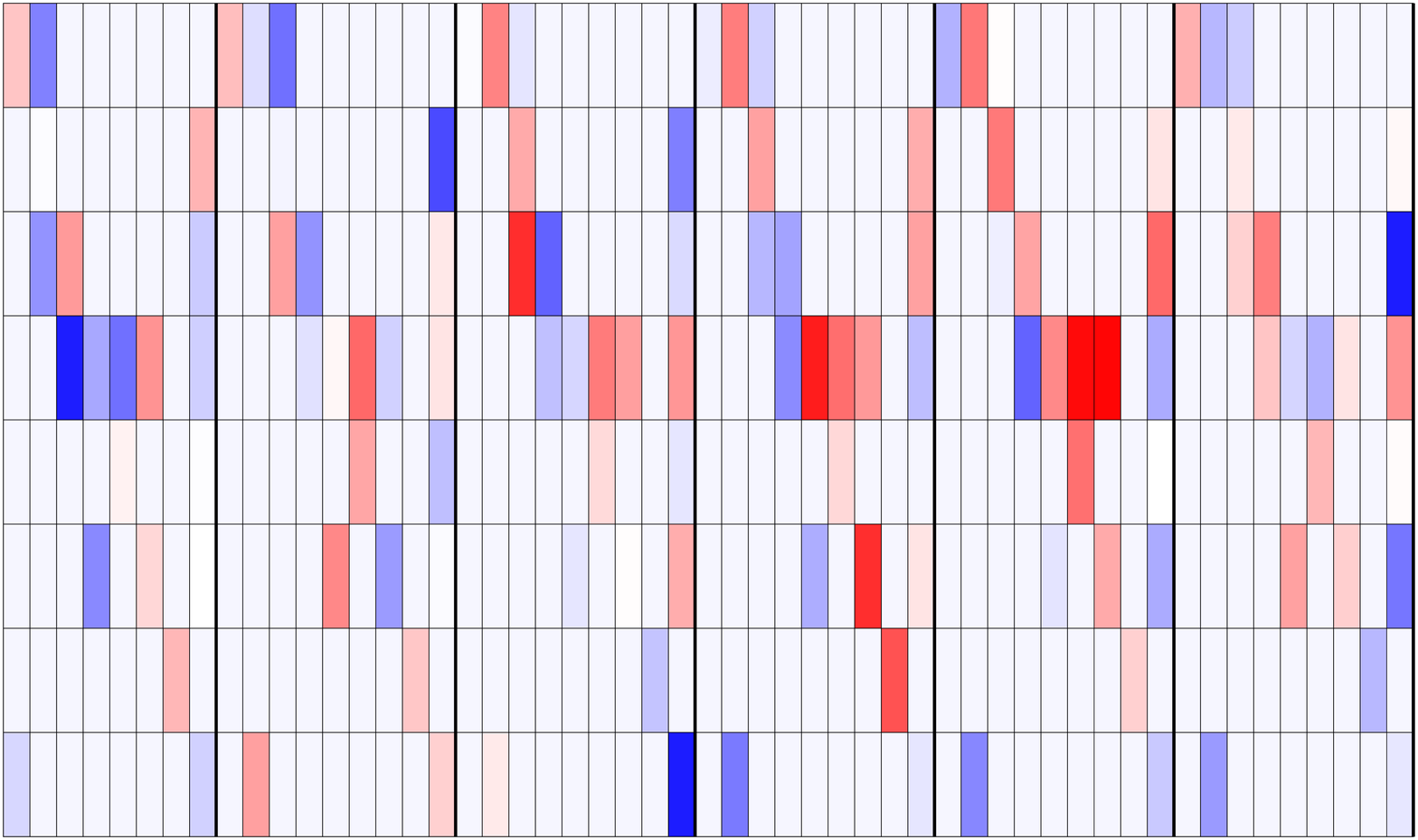}
  \caption{$A00$: the original coefficient matrix used for time-series generation, with sample values $l = 9$ and $p = 6.$, after application of a random mask.}
  \label{fig:a00}
\end{figure}

  \item \textbf{Volume conduction model} \label{sec:org187472a}

  \smallskip

  We used \textsc{FieldTrip} toolbox
  \cite{2011_Oostenveld_FieldTripOpenSourceSoftwareforAdvancedAnalysisofMEGEEGandInvasiveElectrophysiologicalData}
  to generate volume conduction model (VCM) and lead-field matrices.
  VCM was prepared using \textsc{FieldTrip} \newline\lstinline!ft_prepare_headmodel! function
  implementing \lstinline!DIPOLI! method \cite{FieldTripDipoli1989}, 
  which takes as arguments three triangulated surface meshes
  representing the outer surfaces of brain, skull and scalp
  \lstinline!supFunSim/mat/msh.mat! \cite{Brainstorm2011}. VCM is
  available in our toolbox in the precomputed form
  (\lstinline!supFunSim/mat/sel_vol.mat!), although, if required,
  \textsc{FieldTrip} toolbox allows for easy computation of custom
  VCMs on the basis of triangulated meshes which can be obtained from
  structural (T1) MRI scans. Thanks to the modular design of our toolbox 
  other VCMs may also be easily imported. 

  The default head geometry is based on the Colin27%
  \footnote{It can be download from \url{http://neuroimage.usc.edu/bst/download.php?file=tempColin27_2012.zip}
  and contains: \texttt{tess\_innerskull.mat},
  \texttt{tess\_outerskull.mat} and \texttt{tess\_head.mat}.}
  \cite{Brainstorm2011,Colin27-1998,Colin27-2006}. However, it can be
  easily substituted at user's discretion by replacing triangulation
  meshes stored in \lstinline!SETUP.sel_msh! (a list of structures
  containing: scalp outer mesh, skull outer mesh and skull inner mesh,
  where the last triangulation represents ``rough'' brain outer mesh).
  Common choices include realistic head models generated on the basis
  of structural MRI scans or spherical models.

\end{enumerate}



\subsection{Brain Signals in the Sensor Space}
\label{sec:orgbe2fb30}
In the sensor space we need to provide positions for the electrodes
(\emph{a.k.a.}\ sensor montage). By default, in our simulations we use
\emph{HydroCel Geodesic Sensor Net} utilizing 128 channels as EEG cap
layout. Other caps can easily be used by substituting content of the
\lstinline!supFunSim/mat/sel_ele.mat! with electrode position
coordinates obtained either from specific EEG cap producer, or from
the standard montages that are available with EEG analysis software such
as EEGLAB
\cite{2004_Delorme_EEGLABanopensourcetoolboxforanalysisofsingletrialEEGdynamicsincludingindependentcomponentanalysis}
or \textsc{FieldTrip}
\cite{2011_Oostenveld_FieldTripOpenSourceSoftwareforAdvancedAnalysisofMEGEEGandInvasiveElectrophysiologicalData}.
Additionally, for real data acquisition setup, the electrode positions
can be captured using specialized tracking system for every EEG
session.

The volume conduction model, together with source locations and their
orientations, are obtained as described in the three points of the
previous section. Together with electrode positions, they are the
input arguments for the \lstinline!ft_prepare_leadfield! function, 
which is run three times during simulations, outputting $H$,
$H_\mathfrak{i}$ and $H_\mathfrak{b}$.

\section{Implementation Details}\label{implementation-details}


\subsection{Object-oriented approach}\label{object-oriented-approach}

The object-oriented approach provides the toolbox with several
desirable properties of the code and avoids drawbacks of standard
procedural approach commonly employed in \textsc{Matlab} scripts. For
example, \textsc{Matlab} by default stores all variables in one common
workspace. This causes bugs in the code that may be hard to detect. On the other
hand, the object-oriented approach circumvents this difficulty by its
inherent encapsulation property, enclosing variables within a class, 
and sealing it securely from the outside environment. We also note
that the construction of the \textsc{Matlab} language requires
explicit assignment of an instance of the class each time a method
acts on it. This approach necessitates language constructs such as
\lstinline!obj = obj.method!, where \lstinline!obj! is a given
instance of a class.

Data structures created during simulation can be accessed
interactively in the \textsc{Jupyter} notebook or in the 
\textsc{Matlab} script. In particular, property \lstinline!MODEL! from
\lstinline!EEGReconstruction! class contains information about all variables
used within the simulation pipeline.


\subsection{Benefits of literate programming}\label{metaprogramming-and-generation-of-programming-environment}

\begin{figure}[htbp]
  \centering
  \fbox{\includegraphics[width=\textwidth]{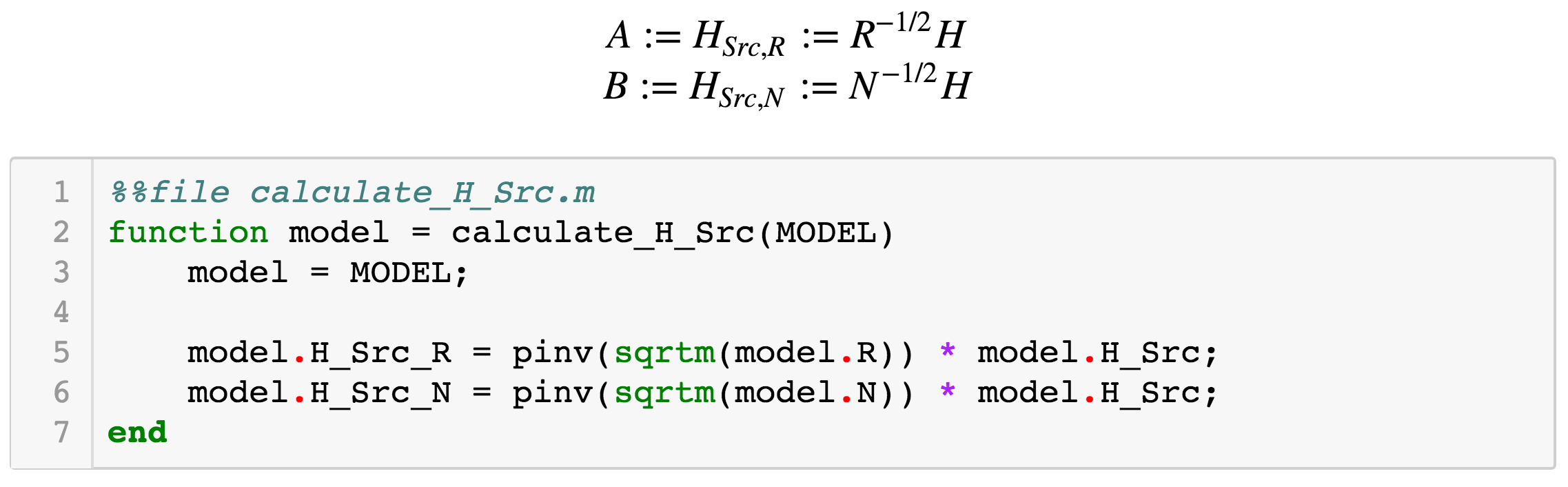}}
  \caption{Example of \textsc{Jupyter} notebook.}
  \label{fig:equat}
\end{figure}

The code of the toolbox was written in \textsc{Jupyter}, which is an
open source application that allows users to create interactive and
shareable notebooks. \textsc{Jupyter} allows for easy export of whole documents to HTML,
\LaTeX, PDF and other formats, and is a very convenient tool for
academic prototyping, because it permits comments in the code 
using \LaTeX\ mathematical expressions. The
source code blocks are interspersed with ordinary natural language
blocks that provide explanations and some insights explaining the 
intrinsic mechanics of the code. Such an approach is
called {\em literate programming} \cite{knuth1984literate}. An example
of mathematical comment and corresponding code is given in
Figure~\ref{fig:equat}.

However, the use of \textsc{Jupyter} is not necessary to run the toolbox.
Instead, the code can be executed under powerful and cross-platform
\textsc{Matlab} environment. To that end we have prepared a version of the
toolbox as the set of \textsc{Matlab} files stored in
\lstinline!supFunSim.zip! archive.

The toolbox does not have a GUI (Graphical User Interface). Instead,
user interacts with it using provided functions. Therefore, as a
prerequisite to use the \textsc{supFunSim} toolbox knowledge of 
\textsc{Matlab} language basics is required.

\subsection{Installation}\label{installation}

Installation of the \textsc{supFunSim} is independent of the operating system. 
For a simple installation similar to \textsc{FieldTrip}'s installation
process, the user can download the file \lstinline!supFunSim.zip! from
{\footnotesize\url{https://github.com/nikadon/supFunSim.git}}. This archive contains the
whole toolbox. After unpacking this archive, the user should execute
\lstinline!addpath(genpath('/path/to/toolboxes/supFunSim/']))!.
Function \lstinline!genpath! will ensure that all subdirectories will
be added to your path. It is most convenient to have the
\lstinline!addpath! function in the \lstinline!startup.m! script
located in the \textsc{Matlab} directory. Then, the user may run the
\lstinline!RunAll.m! script (preferably line by line, in order to
follow execution). The user has to make sure that there is a
\lstinline!mat/! directory (or a link to it) containing
\lstinline!mat! files required by the toolbox in the toolbox
directory. The \lstinline!mat! files are available for download at
{\footnotesize\url{http://fizyka.umk.pl/~tpiotrowski/supFunSim}}.

More advanced user may manipulate {\sc Jupyter}
notebooks directly and use \lstinline!make! tool to
set up the toolbox from scratch. Namely, in order to open and run
notebooks the user should download and install \textsc{Jupyter} notebook
with \textsc{Matlab} kernel. The easiest way to do it (under a
\textsc{Unix}-like system) is by executing the following instructions
in the command line. First, we set up a virtual environment, which
will install \textsc{Python} packages locally:

\begin{lstlisting}[language=sh]
sudo pip install virtualenv # installing virtualenv environment
mkdir supFunSimToolbox # making directory for virtual environment
unzip supFunSim.zip -d supFunSimToolbox # extracting toolbox
virtualenv supFunSimToolbox # creating virtual environment
source ./bin/activate # activating virtual environment
\end{lstlisting}

Next, we install all necessary packages and install \textsc{Matlab}
Engine API for \textsc{Python} 

\begin{lstlisting}[language=sh]
pip install -r requirements.txt # installing all requirements
cd /path/to/matlabroot/extern/engines/python
python setup.py install
\end{lstlisting}

We also provide a \lstinline!make! tool for a simple administration of
notebooks' code. For example, the user may execute \lstinline!make!
\lstinline!everything! in terminal in order to generate all source
code files. See \lstinline|README.md| file in the repository for
details.

At this stage, one can run the simulations and ``play with'' the code
by going to \lstinline!supFunSimToolbox! directory and running

\begin{lstlisting}[language=sh]
jupyter-notebook
\end{lstlisting}

Finally, the description of the installation under Windows can be
found in the \lstinline|README.md| file.

\subsection{Prerequisites/dependencies}\label{PrerequisitesDependencies}

In addition to \textsc{Matlab} our toolbox requires 3 packages: 
\textsc{FieldTrip} toolbox (version at least 20150227) \cite{FieldTrip2011},
\textsc{MVARICA} toolbox (version at least 20080323) \cite{gomez2008measuring},
\textsc{ARfit} toolbox (version at least 20060713) \cite{2001_Neumaier_Estimationofparametersandeigenmodesofmultivariateautoregressivemodels,2001_Schneider_Algorithm808ARfitamatlabpackagefortheestimationofparametersandeigenmodesofmultivariateautoregressivemodels}.
Location of these toolboxes should be added to \textsc{Matlab} \texttt{path}.

%

\section{Application structure}\label{application-structure}

The simulation framework provided with the current paper consists of a
set of modules represented by corresponding classes. The classes are
defined in separate (self-contained) notebooks. The classes depend on
auxiliary functions generated alongside with them when appropriate
\lstinline!make! target is invoked. In this way, a given class is
enclosed and all operations involving it are made within it. The
toolbox contains six classes (described in the next section) with a
number of auxiliary functions. 

\begin{figure}[htbp]
  \centering
  \includegraphics[width=9cm]{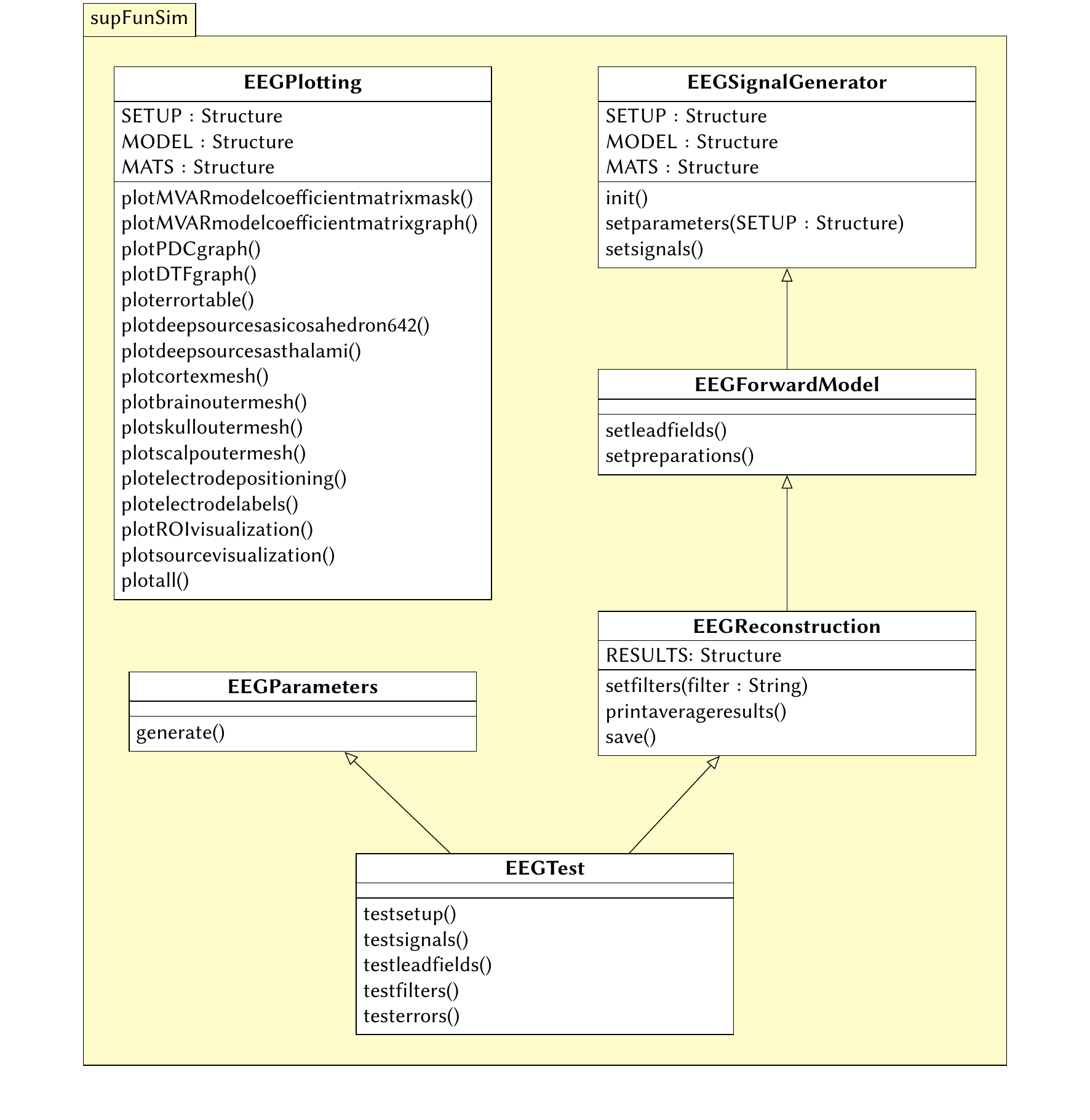}
  \caption{Dependencies between classes.}
  \label{fig:classes}
\end{figure}

\subsection{Overview of toolbox classes}
The main functionality of the toolbox is provided by the following
five classes:

\begin{itemize}
\itemsep1pt\parskip0pt\parsep0pt
\item
  \lstinline!EEGParameters.ipynb! --- class generating parameters for
    simulations. It can be overwritten in order to obtain desired
    parameters for a sequence of simulations.
\item
  \lstinline!EEGSignalGenerator.ipynb! --- class used to generate
    signal for forward modelling of sources. It can be
    overwritten to generate a signal with given or desired properties.
\item
  \lstinline!EEGForwardModel.ipynb! --- class implementing forward
    model. It constructs and adds together all signals
    (source activity, background activity and interference noise). Furthermore,
    the lead-field matrix is build using \textsc{FieldTrip} library
    based on the selected head model.
\item
  \lstinline!EEGReconstruction.ipynb! --- class implementing methods
    used in the reconstruction of the underlying neuronal activity.
    All spatial filters are implemented in this class. 
\item
  \lstinline!EEGPlotting.ipynb! --- class implementing plots detailing
    execution of experiments. Various visualizations are accessible
    through the methods included in this class.
\end{itemize}

We also wrote a class for unit testing of the toolbox functionality:

\begin{itemize}
\itemsep1pt\parskip0pt\parsep0pt
\item
  \lstinline!EEGTest.ipynb! --- class implementing unit tests and
    validation of the code against the functional-code toolbox
    implementation.
\end{itemize}

Fig.~\ref{fig:classes} gives an overview of relationships between
implemented classes. We should also emphasize that modular design
facilitates reuse and extensibility of the source code and adaptation to
other applications. For example, if one wishes to generate source
signals in a different way compared with the implemented version, one
needs only to overwrite \lstinline!EEGSignalGenerator! class (or some
of its methods) while keeping the rest of the code and its
functionality intact.

\subsection{\lstinline!Mat! files}

Directory \lstinline!mat/! contains third-party data with the geometry
of the brain, taken from the \textsc{Brainstorm} toolbox \cite{Brainstorm2011} 
and extracted using \textsc{FieldTrip} procedures:

\begin{itemize}
  \itemsep1pt\parskip0pt\parsep0pt
  \item
    \lstinline!sel_msh.mat! --- head compartments geometry
    (vertices and triangulation forming meshes for brain, skull and scalp);
    this data
    can be used as an input for volume conduction model
    and
    lead-field generation using \textsc{FieldTrip}
    (or any other toolbox that can be used to generate forward model);
  \item
    \lstinline!sel_vol.mat! --- volume conduction model (head-model).
    This structure contains head compartments geometry
    (the same as in \lstinline!sel_msh.mat!) accompanied by
    their conductivity values and a
    matrix containing
    numerical solution (utilizing boundary or finite element method)
    to a system of differential equations describing propagation of the electric field.
    This data is obtained using
    \textsc{FieldTrip}'s \lstinline!dipoli! method
    and is used as an input to the function that calculates the lead-field matrix.
    The default volume conduction model was prepared in accordance with
      the instruction provided in the \textsc{FieldTrip} tutorial
      \emph{Creating a BEM volume conduction model of the head for source-reconstruction of EEG data}.%
      \footnote{Available at \url{http://www.fieldtriptoolbox.org/tutorial/headmodel_eeg_fem/}} %
    This structure is used to compute lead-fields;
  \item
    \lstinline!sel_geo_deep_thalami.mat! --- mesh containing candidates
    for location of deep sources (based on \textit{thalami}).
    The mesh was prepared on the basis of the Colin27
      \cite{Brainstorm2011,Colin27-1998,Colin27-2006}
      MRI images;
  \item
    \lstinline!sel_geo_deep_icosahedron642.mat! --- mesh containing
    candidates for location of deep sources (based on
    \textit{icosahedron642});
  \item
    \lstinline!sel_atl.mat! --- cortex geometry with (anatomical) ROI
    parcellation (cortex atlas). This detailed triangulation is parceled into cortical patches
      (a.k.a. regions of interest, ROIs). It
      contains a 15000 vertices and it is based on the sample data
      accompanying the BrainStorm toolbox \cite{Brainstorm2011}.
      It was originally prepared using
      FreeSurfer \cite{FreeSurfer1999,FreeSurferCortex2004} software;
  \item
    \lstinline!sel_ele.mat! --- geometry of electrode positions.
    By default we use \emph{HydroCel Geodesic Sensor Net}
      sensor montage utilizing 128 channels available.
      The electrode positions file is available with the
      \textsc{FieldTrip} toolbox as \texttt{GSN-HydroCel-128.sfp} file;
  \item
    \lstinline!sel_src.mat! --- lead-fields of all possible source locations.
\end{itemize}


\subsubsection{Simulation parameters class}

This class is responsible for setting up parameters of simulations.

\begin{itemize}
\itemsep1pt\parskip0pt\parsep0pt
\item
  \lstinline!EEGParameters!:

  \begin{itemize}
  \itemsep1pt\parskip0pt\parsep0pt
  \item
    \lstinline!generate! --- this method generates the set of parameters
      for simulations. The method itself is mainly
      based on \lstinline!generatedummysetup! function which itself uses \lstinline!setinitialvalues!
      and \lstinline!setsnrvalues! functions containing default configuration
      for the reconstruction. Users willing to change basic
      configuration should edit the 
      \lstinline!configurationparameters.m! file. The assignment of parameters' values 
      made in this file overwrites default parameter settings.

      For the complete list of all simulation parameters consult Table~\ref{tab:setup}.

      For unit testing, configuration from the \lstinline!testparameters! should be used. The configuration in this file agrees with the configuration used in the 
      \lstinline!supFunSim.org! file. To perform unit testing, simply uncomment appropriate
      line in the \lstinline!generatedummysetup.m! file.
  \end{itemize}
\end{itemize}

\subsubsection{Signal generation class}\label{signal-generation}

This class is responsible for EEG signal generation.

\begin{itemize}
\itemsep1pt\parskip0pt\parsep0pt
\item
  \lstinline!EEGSignalGenerator!:

  \begin{itemize}
  \itemsep1pt\parskip0pt\parsep0pt
  \item
    \lstinline!init! --- this method initializes all toolboxes required by \textsc{supFunSim}.
    It sets path to toolboxes, creates their default settings, sets
      head model, geometry of patches etc.;
  \item
    \lstinline!setparameters! --- this method sets configuration for simulation
      using parameters from \lstinline!EEGParameters! class;
  \item
    \lstinline!setsignals! --- this method generates source-level signals: 
      \lstinline!sim_sig_SrcActiv.sigSRC!, \lstinline!sim_sig_IntNoise.sigSRC!,
      \lstinline!sim_sig_BcgNoise.sigSRC!, as well as sensor level \lstinline!sim_sig_MesNoise.sigSNS! measurement noise:
  \begin{itemize}
    \itemsep1pt\parskip0pt\parsep0pt
    \item
      \lstinline!makeSimSig! ---  MVAR-based signal generation; the signal is then divided into \emph{pre} and \emph{pst} parts,
    \item \lstinline!generatetimeseriessourceactivity! --- generates signals of sources of interest \lstinline!sim_sig_Src_Activ.sigSRC! using \lstinline!makeSimSig! and if required, adds ERP deterministic signal to the \emph{pst} part of the signal of interest,
    \item \lstinline!generatetimeseriesinterferencenoise! --- generates interference noise signal \lstinline!sim_sig_IntNoise.sigSRC! as the negative of \lstinline!sim_sig_SrcActiv.sigSRC! signal of interest with added white Gaussian noise of prescribed power relative to the power of \lstinline!sim_sig_SrcActiv.sigSRC!,
            \item \lstinline!generatetimeseriesbackgroundnoise! --- generates background activity signal \lstinline!sim_sig_BcgNoise.sigSRC! using \lstinline!makeSimSig!,
      \item \lstinline!generatetimeseriesmeasurementnoise! --- generates measurement (sensor-level) noise signal \lstinline!sim_sig_MesNoise.sigSNS! as an additive white Gaussian noise.
  \end{itemize}    
  \end{itemize}
\end{itemize}

\subsubsection{Forward model class}\label{forward-model}

Class \lstinline!EEGForwardModel! generates solution to the
forward problem: leadfield matrices and the resulting electrode-level signal.

\begin{itemize}
\itemsep1pt\parskip0pt\parsep0pt
\item
  \lstinline!EEGForwardModel!:

  \begin{itemize}
  \itemsep1pt\parskip0pt\parsep0pt
  \item
    \lstinline!setleadfields! --- this method generates leadfield matrices:

    \begin{itemize}
    \itemsep1pt\parskip0pt\parsep0pt
    \item
      \lstinline!geometryrandomsampling! --- random or user-defined selection of cortex ROIs
      for sources (of interest, interfering activity, background activity),
    \item
      \lstinline!geometryindices! --- identification of cortex ROIs' indices within cortex atlas,
    \item
      \lstinline!geometrycoordinates! --- coordinates of vertices of sources within selected ROIs,
    \item
      \lstinline!geometrydeepsources! --- coordinates of vertices of sources within thalamus,
    \item
      \lstinline!geometryperturbation! --- generation of perturbed source locations and orientations, 
    \item
      \lstinline!geometryleadfieldscomputation! --- computation of original lead-fields of sources of interest \lstinline!sim_lfg_SrcActiv_orig.LFG!, sources of interfering activity \lstinline!sim_lfg_IntNoise_orig.LFG!, sources of background activity \lstinline!sim_lfg_BcgNoise_orig.LFG!, as well as their respecitve perturbed versions \lstinline!sim_lfg_SrcActiv_pert.LFG!, \lstinline!sim_lfg_IntNoise_pert.LFG!, and\linebreak \lstinline!sim_lfg_BcgNoise_pert.LFG!, respectively,
    \item
      \lstinline!forwardmodeling! --- multiplication of source signals by their corresponding lead-field matrices yielding sensor-level signals of sources of interest \lstinline!sim_sig_SrcActiv.sigSNS!, sources of interfering activity \lstinline!sim_sig_IntNoise.sigSNS!, and sources of background activity\linebreak \lstinline!sim_sig_BcgNoise.sigSNS!;
    \end{itemize}
  \item
    \lstinline!setpreparations! --- generates output EEG signal and prepares for signal reconstruction using spatial filters:
    
    \begin{itemize}
    \itemsep1pt\parskip0pt\parsep0pt
    \item
      \lstinline!preparationsnrsadjustment! --- rescales sensor-level signals
      to the appropriate SNRs,
    \item
      \lstinline!prepareleadfieldconfiguration! --- determines whether original or perturbed lead-field matrices of sources of interest and of interfering sources will be made available to spatial filters according to the user's setting of \lstinline!SETUP.H_Src_pert! and \lstinline!SETUP.H_Int_pert! flags; moreover, reduces the rank of lead-field matrix of interfering sources according to \lstinline!SETUP.IntLfgRANK! variable value,
    \item
      \lstinline!preparemeasuredsignals! --- sums sensor-level signals and adds measurement noise signal (\lstinline!sim_sig_MesNoise.sigSNS!) to produce output \lstinline!y_Pre! and \lstinline!y_Pst! EEG signals.
    \item
      \lstinline!store2eeglab! --- a method that allows the user to save data in the EEGLAB format; this toolbox and its plugins allows for better preprocessing and visualization of EEG signal.
    \item \lstinline!rawAdjTotSNRdB! --- adjusts \lstinline!sim_sig_IntNoise.sigSNS!, \lstinline!sim_sig_BcgNoise.sigSNS! and \lstinline!sim_sig_MesNoise.sigSNS! signals' power with respect to the power of 
\lstinline!sim_sig_SrcActiv.sigSNS!, to obtained desired signal-to-interference, signal-to-background-activity, and signal-to-noise ratios, respectively. These ratios are defined using \lstinline!SETUP.SINR!, \lstinline!SETUP.SBNR!, and \lstinline!SETUP.SMNR! variables, respectively, and are expressed in decibels [dB] using the following implementation:
\begin{lstlisting}[language=sh]
function [y] = rawAdjTotSNRdB(x01, x02, newSNR)
  y = ((x02 / norm(x02)) * norm(x01)) / (db2pow(0.5 * newSNR)),
end
\end{lstlisting}
where \lstinline!db2pow! is the \textsc{Matlab} function
converting decibels to power.
    \end{itemize}
  \end{itemize}
\end{itemize}

\subsubsection{Reconstruction class}\label{reconstruction}
Class \lstinline!EEGReconstruction! computes implemented spatial filters and applies them to the observed \lstinline!y_Pst! simulated EEG signal. It also computes various fidelity measures of the reconstruced activity of sources of interest.

\begin{itemize}
\itemsep1pt\parskip0pt\parsep0pt
\item
  \lstinline!EEGReconstruction!:

  \begin{itemize}
  \itemsep1pt\parskip0pt\parsep0pt
  \item \lstinline!setfilters! --- calculates matrices which are
    used in the process of reconstruction:

    \begin{itemize}
    \itemsep1pt\parskip0pt\parsep0pt
    \item
      \lstinline!spatialfilterconstants! --- We compute some of
        constants used later in defining filters. 
    \item
      \lstinline!spatialfiltering! --- We compute all intermediate
        variables needed to calculate the filters. 
    \item
      \lstinline!spatialfilteringexecution! --- For every filter
        $W(\theta)$ given as a parameter to this function we are
        calculating $W(\theta) y$ using post-interval signal as $y$.
        Then we perform \lstinline!arfit! for all reconstructed
        signals and obtain autoregression matrix. This matrix is
        necessary to calculate PDC and DTF measures.
    \item
      \lstinline!spatialfilteringerrorevaluation! --- Function
        which calculates difference between original and
        reconstructed signals. It uses various measures: Euclidean
        metric and correlation coefficients to compare activity
        signals, MVAR coefficient matrices and PDC and DTF coefficient
        matrices.
    \item
      \lstinline!vectorizerrorevaluation! --- Function that combines
        results in a single array. Such uniform output of different
        error measures is later used in plotting.
    \end{itemize}
  \item
    \lstinline!printaverageresults! --- print table of comparison of
      different reconstruction filters.
  \item
    \lstinline!save! --- save reconstructed filters.
  \end{itemize}
\end{itemize}

\subsection{Plotting class}\label{plotting}

The toolbox makes it possible to visualize results of experiments.
User can plot results of simulation using \lstinline!EEGPlotting!
class which is specially prepared for this.

\begin{lstlisting}[language=Matlab]
eegplot = EEGPlotting(reconstruction);
\end{lstlisting}

\begin{figure*}[t!]
    \centering
        \centering
        \includegraphics[width=\textwidth]{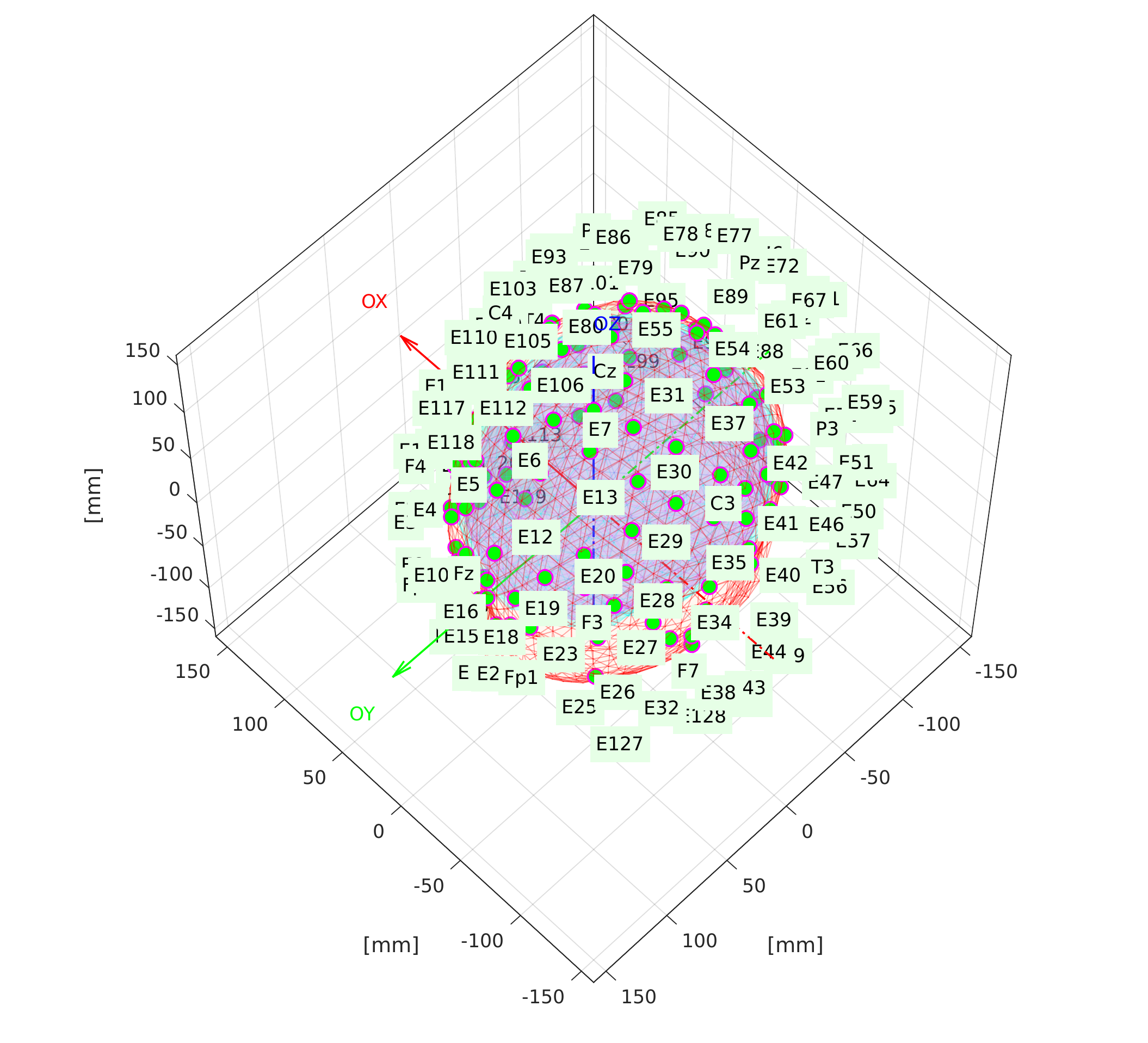}
    \caption{Volume conduction model essential components. Triangulation meshes
   representing brain, skull and scalp boundaries with electrode
   positions plotted on top of the scalp surface.
   This figure was generated using an instance of
   \textbf{\texttt{EEGPlotting}}
   class
   employing
   \texttt{plotbrainoutermesh()},
   \texttt{plotskulloutermesh()},
   \texttt{plotscalpoutermesh()},
   \texttt{plotelectrodepositioning()}
   and
   \texttt{plotelectrodelabels()}
   methods.}
    \label{fig:viz1}
\end{figure*}

\begin{figure*}[t!]
    \centering
        \centering
        \includegraphics[width=\textwidth]{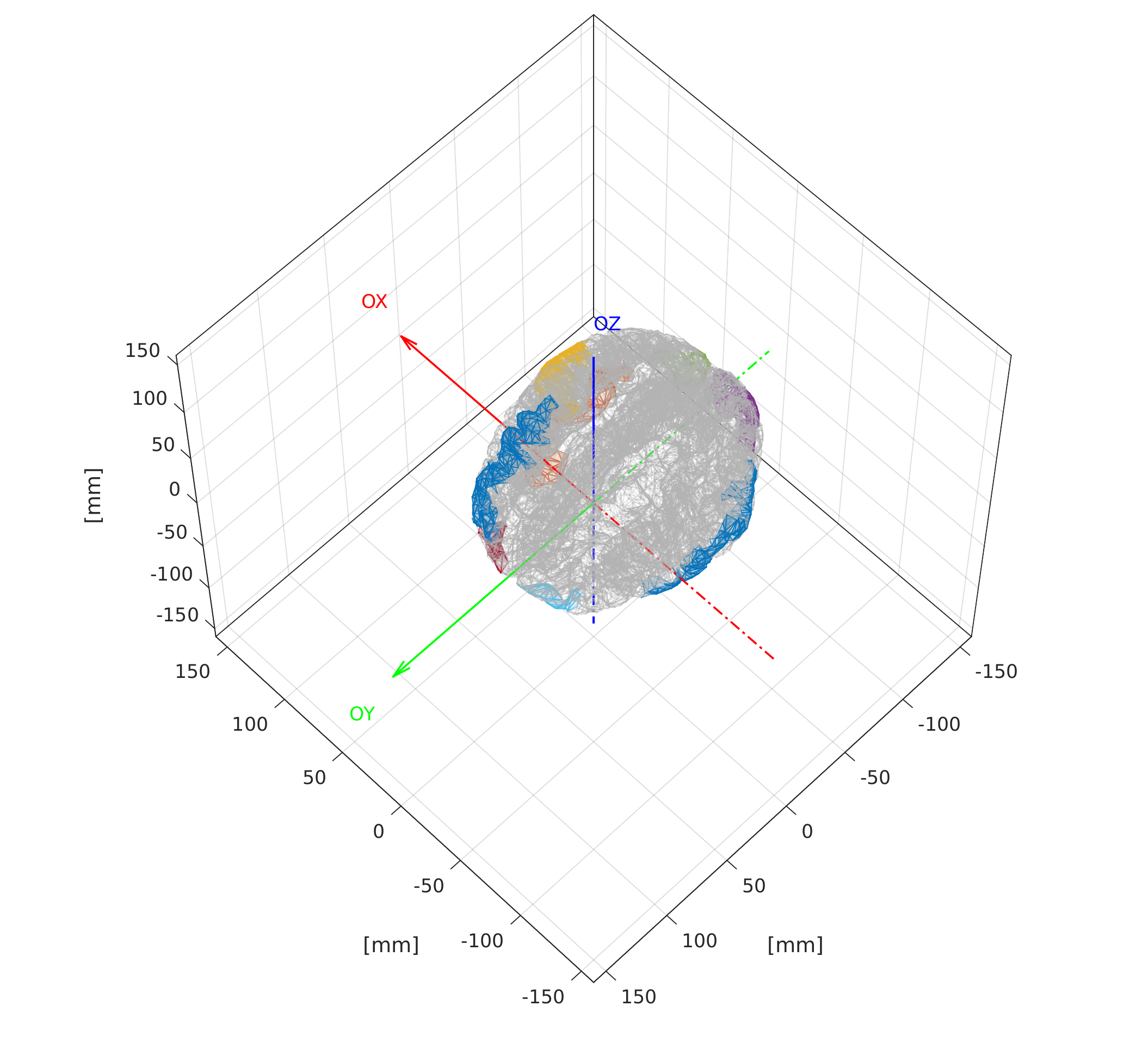}
    \caption{Cortex and ROIs. Detailed cortical surface triangulation
   with selected cortical patches.
   This figure was generated using an instance of
   \textbf{\texttt{EEGPlotting}}
   class
   employing
   texttt{plotcortexmesh()}
   and
   texttt{plotROIvisualization()}
   methods.}
    \label{fig:viz2}
\end{figure*}

\begin{figure*}[t!]
    \centering
        \centering
        \includegraphics[width=\textwidth]{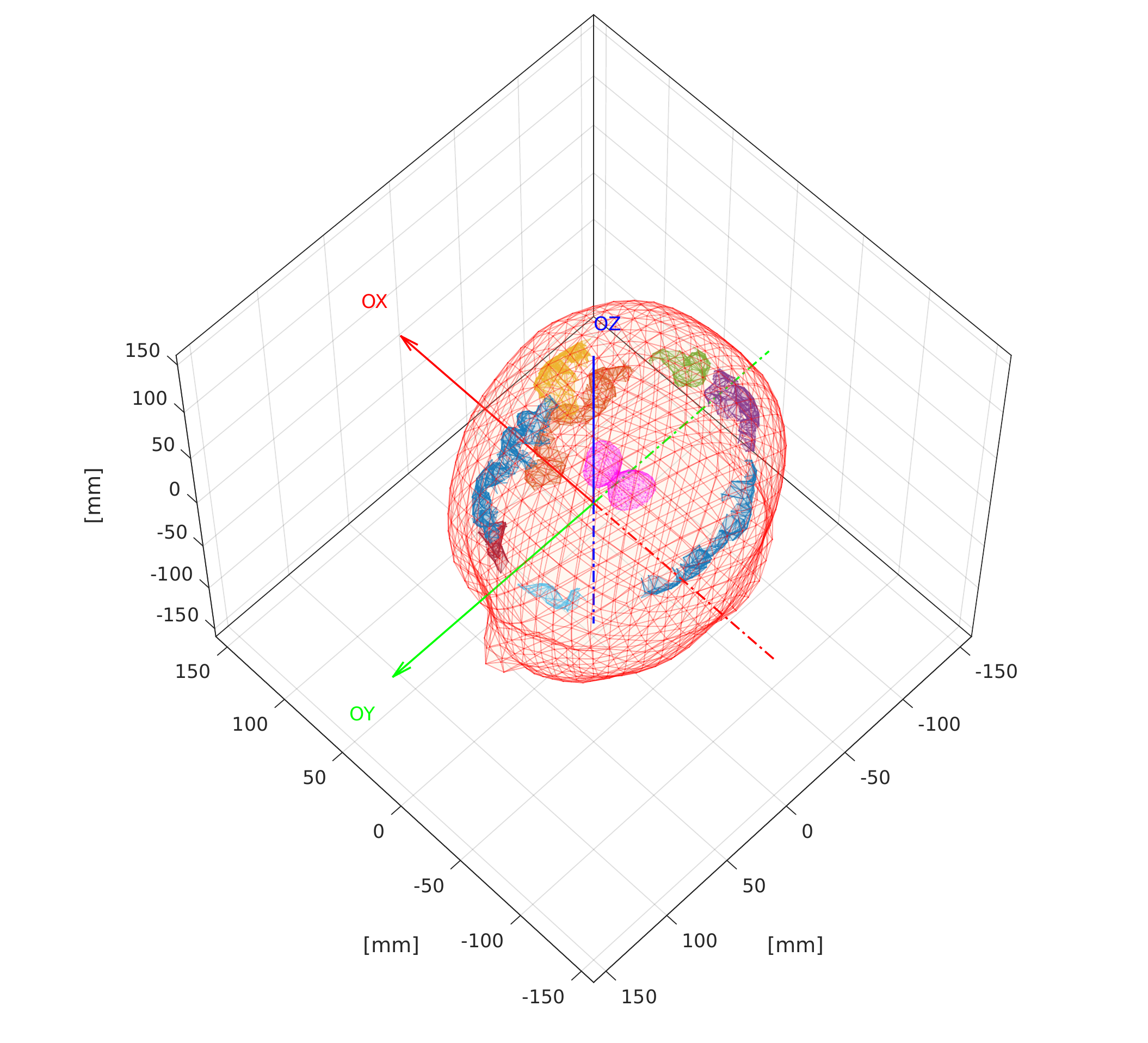}
    \caption{ROIs and thalami.
   Cortical patches selected as a candidate ROIs for source position
   with thalami mesh and scalp outer mesh.
   This figure was generated using an instance of
   \textbf{\texttt{EEGPlotting}}
   class
   employing
   \texttt{plotROIvisualization()},
   \texttt{plotdeepsourcesasthalami()}
   and
   \texttt{plotscalpoutermesh()}
   methods.}
    \label{fig:viz3}
\end{figure*}

\begin{figure*}[t!]
    \centering
        \centering
        \includegraphics[width=\textwidth]{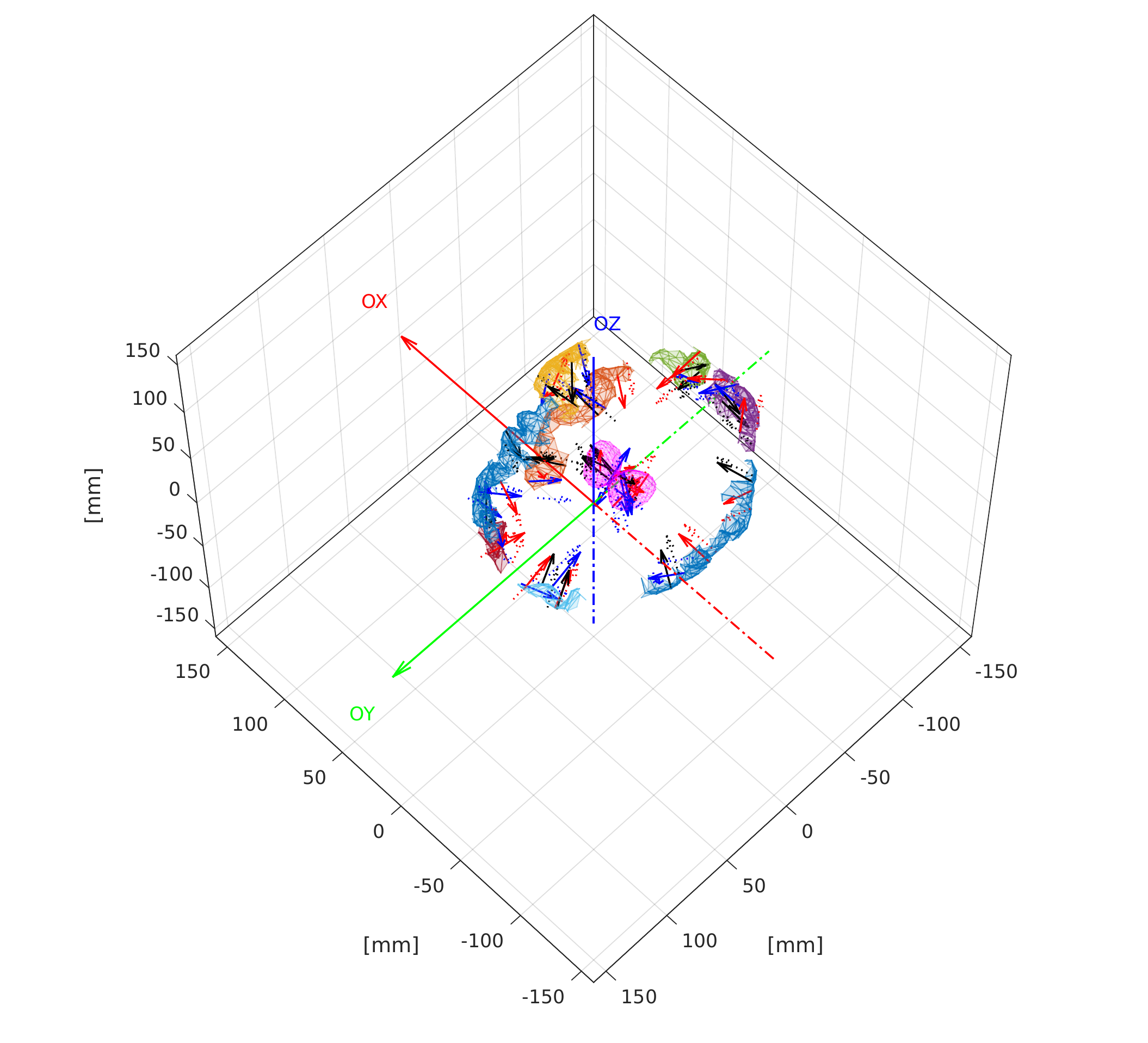}
    \caption{Bioelectrical activity positions and orientations.
   Cortical patches selected as a candidate ROIs for source position
   with thalami mesh. Vectors representing direction of the dipole
   moments for the sources of bio-electrical activity. Red lines
   represent the activity of interest; blue -- the interfering
   activity and black -- background activity.
   Solid lines represent original sources
   and dotted lines represent perturbed sources.
   Arrows representing
   dipole position and orientation are drawn not to scale.
   This figure was generated using an instance of
   \textbf{\texttt{EEGPlotting}}
   class
   employing
   \texttt{plotROIvisualization()},
   \texttt{plotdeepsourcesasthalami()}
   and
   \texttt{plotsourcevisualization()}
   methods.}
    \label{fig:viz4}
\end{figure*}

Interesingly, there are many ways to facilitates visualization of the
analysis and results. \textsc{Jupyter} functionality gives us a
possibility to plot figures inside notebook (using \lstinline!%plot! magic
option):

\begin{lstlisting}[language=Matlab]
%plot inline
eegplot.plotskulloutermesh();
\end{lstlisting}

or to open it in \textsc{Matlab} interactive environment:

\begin{lstlisting}[language=Matlab]
%plot native
eegplot.plotsourcevisualization();
\end{lstlisting}

Toolbox \textsc{supFunSim} contains variety of plotting functions for
different visualizations of results. Some of them are presented in
Figures~\ref{fig:viz1}~\ref{fig:viz2}~\ref{fig:viz3} and \ref{fig:viz4}.

Plot consists of layers that are generated by functions with
self-explonatory names. E.g. function \lstinline!plotROIvisualization!
plots cortex, regions of interest. Function
\lstinline!plotsourcevisualization! plots mesh for ROIs on cortex,
mesh for deep sources ROI, sources and cortex.

\begin{itemize}
\itemsep1pt\parskip0pt\parsep0pt
\item
  \lstinline!EEGPlotting!:

  \begin{itemize}
  \itemsep1pt\parskip0pt\parsep0pt
  \item
    \lstinline!plotMVARmodelcoefficientmatrixmask! --- this method plots mask for MVAR model coefficient matrix.
  \item
    \lstinline!plotPDCgraph! --- this method is used for plotting PDC profiles across sources of interest and interfering sources.
  \item
    \lstinline!plotDTFgraph! --- this method is used for plotting DTF profiles across sources of interest and interfering sources.
  \item
    \lstinline!plotMVARmodelcoefficientmatrixgraph! --- this method plots the matrix of composite MVAR model for sources of interest, interfering and background sources.
  \item
    \lstinline!ploterrortable! --- this method plots results of reconstruction as heatmap table.
  \item
    \lstinline!plotdeepsourcesasicosahedron642! --- this method is for plotting of deep sources.
  \item
    \lstinline!plotdeepsourcesasthalami! --- this method can be used for plotting of both thalami.
  \item
    \lstinline!plotcortexmesh! --- this method plots cortex mesh.
  \item
    \lstinline!plotbrainoutermesh! --- this method plots brain outer mesh.
  \item
    \lstinline!plotskulloutermesh! --- this method plots skull outer mesh.
  \item
    \lstinline!plotscalpoutermesh! --- this method plots scalp outer mesh.
  \item
    \lstinline!plotelectrodepositioning! --- this method plots electrode positions.
  \item
    \lstinline!plotelectrodelabels! --- this method plots electrode labels.
  \item
    \lstinline!plotROIvisualization! --- this method plots ROIs based on generated meshes.
  \item
    \lstinline!plotsourcevisualization! --- this method is used for source visualization.
  \end{itemize}
\end{itemize}

\subsection{Unit test class}\label{unit-test}

Since this implementation is based on the previous one, which was done
in \lstinline!Org-mode!, authors have created a class
\lstinline!EEGTest! for unit tests.

In order to generate and distribute files into directories (necessary
for the test) use a \lstinline!make! target \lstinline!test!.

For example, unit tests can look like that:

\begin{lstlisting}[language=Matlab]
eegtest = EEGTest()
eegtest = eegtest.testsetup()
eegtest = eegtest.testsignals()
eegtest = eegtest.testleadfields()
eegtest = eegtest.testfilters()
eegtest = eegtest.testerrors()
\end{lstlisting}

This class can be useful for developers who wants to extend our
framework. This way they always can check whether it still passes
the compliance test.

\section{Sample usage}\label{sample-usage}

\subsection{Generating set of parameters for simulations}\label{generating-set-of-parameters-for-simulations}

Generation of a new simulation requires preparation of a set of parameters. 
This is done by the \lstinline!EEGParameters!  class, in which the \lstinline!generate!
method is included. In the default version the \lstinline!dummy!
function is called, which returns the default parameter structure, but
the user can always overwrite it to create her/his own version, 
or change the configuration of the simulation and perform own
experiments.  Syntax for generating parameters is as follows: 

\begin{lstlisting}[language=Matlab]
parameters = EEGParameters().generate();
\end{lstlisting}

Several sample settings have been prepared. Functions
\lstinline!setinitialvalues! and \lstinline!setsnrvalues! will set up
initial parameters and signal to noise ratios. 
Function \lstinline!smartparameters! overwrites initial values and contains
parameters for the sample run. 
Function \lstinline!testparameters! contains parameters for unit test 
done by \lstinline!EEGTest!. 
To use it line containing this function must be uncommented.

Generated structure contains about 50 fields. 
All fields are listed in Table~\ref{tab:setup}.

\subsection{Example run of simulations}\label{example-run-of-simulations}

The input configuration structure from \lstinline!EEGParameters! class
contains options and parameters that specify how the stimulation will
run. Once the user is satisfied with the parameter settings a sequence 
of simulations is ready to run, which may look, e.g., like that:

\begin{lstlisting}[language=Matlab]
filters = [ 'LCMV', 'MMSE', 'ZF', 'RANDN', 'sMVP_R' ];
reconstruction = EEGReconstruction();
reconstruction = reconstruction.init();
for np = 1:length(parameters)
  parameter = parameters(np)
  reconstruction = reconstruction.setparameters(parameter);
  reconstruction = reconstruction.setsignals();
  reconstruction = reconstruction.setleadfields();
  reconstruction = reconstruction.setpreparations();
  reconstruction = reconstruction.setfilters(filters);
  reconstruction.save();
end
reconstruction = reconstruction.printaverageresults();
\end{lstlisting}

Here \lstinline!parameters! were generated as described above. 
Selection of filters to compute the source reconstruction is done above 
in a very direct way. Filters are self-contained, i.e. they can run independently. 
What is worth noting, there are fifteen spatial filters
available in the current version of \textsc{supFunSim} (including, e.g.,
classical LCMV), and more will be added.


All intermediate values of model variables along with the initial
settings are stored in attributes \lstinline!SETUP! and
\lstinline!MODEL!. Attribute \lstinline!RESULTS! contains the scores 
of all the most important measures of errors for individual filters.
Moreover, all meshes are kept in attribute \lstinline!MATS!. The main
reason is that meshes are loaded only once during all iterations. All
that, as a part of the main object, can be saved in \lstinline!mat!
file and later restored.

\begin{lstlisting}[language=Matlab]
load('reconstruction_DATE.mat');
obj.MODEL
\end{lstlisting}
Here \lstinline!DATE! is identifier of reconstruction, given by date of execution.

\section{Conclusion}\label{conclusions}

Many tools have been created for EEG/MEG source reconstruction and localization. To the best 
of our knowledge our \textsc{supFunSim} toolbox is the first written in modular and object-oriented way that 
allows to use many spatial filters for EEG source reconstruction. Except for classical 
linearly constrained minimum-variance (LCMV) filter, 
\cite{Frost1972,VanVeen1997,Sekihara2008}, eigenspace LCMV \cite{Sekihara2008}, nulling (NL)
\cite{Hui2010}, it also contains our new minimum-variance pseudo-unbiased reduced-rank
(MV-PURE) filters \cite{Piotrowski2008,Piotrowski2009,Piotrowski2019}. 
Our toolbox may thus enable an easy comparison of accuracy of different filters.   
It may be used in combination with the standard \textsc{FieldTrip} or \textsc{EEGLab} packages, and should be useful in construction of both forward and inverse models. The use of reconstructed signals for the directed connectivity analysis using partial directed coherence (PDC)  \cite{Baccala2001} and directed transfer function (DTF) \cite{Kaminski2001} measures is also of great interest in analysis of information flow in the brain. 

Future work will proceed on the further development of  \textsc{supFunSim} toolbox, and on the testing of the methods on real data from EEG experiments. Many large-scale subnetworks have been discovered using fMRI, and a big challenge is to characterize their dynamics. EEG signals after source reconstruction may help to analyze fast switching between different subnetworks, discovering fingerprints of brain's cognitive activity.      

\textsc{Matlab} is still a very popular language among neuroscientists and it was 
an obvious choice due to availability of functions that could be adopted from other toolboxes.
However, we plan to make a \textsc{Python} implementation and thorough
simulations illustrating the power of our toolbox, comparing different filters. 
The code in the current form that contains objects should be easily rewritten to \textsc{Python}.

\appendix
\section{Appendices} \label{sec:appendices}

\subsection{Implemented spatial filters} \label{sec:filters}
We denote the concatenated composite lead-field matrices of $H$ and
$H_\mathfrak{i}$ as $H_\mathfrak{c} := [H\ H_\mathfrak{i}]$, and
similarly $q_\mathfrak{c} := [q^t q_\mathfrak{i}^t]^t.$ The covariance
matrices of $q$, $q_\mathfrak{c}$, $q_\mathfrak{b} + n_\mathfrak{m}$,
$y$ are denoted by $Q$, $Q_\mathfrak{c}$, $N$, $Y$, respectively.

We selected for comparison the following spatial filters:
\begin{enumerate}
  \item The LCMV filter, expressed as both~\cite{Moiseev2011}
    \begin{equation}
      W_{LCMV(R)} := (H^t R^{-1} H)^{-1} H^t R^{-1},
    \end{equation}
    and~\cite{Moiseev2015}
    \begin{equation}
      W_{LCMV(N)} := (H^t N^{-1} H)^{-1} H^t N^{-1}.
    \end{equation}
  \item The nulling filter~\cite{Hui2010}:
    \begin{equation}
      W_{NL} := [\mathbb{I}_l\ 0_{l\times k}] (H_c^t R^{-1} H_c)^{-1} H_c^t R^{-1}.
    \end{equation}
  \item The Wiener filter, defined as~\cite{Kailath2000}
    \begin{equation}
      W_{F{\mhyphen}MMSE} := QH^tR^{-1},
    \end{equation}
    for the interference-free model, and~\cite{Kailath2000}
    \begin{equation}
      W_{I{\mhyphen}MMSE} := \mathbb{E}[qq_c^t] H_c^t R^{-1},
    \end{equation}
    for the model in presence of interference.
  \item The zero-forcing filter, defined as~\cite{Kailath2000}
    \begin{equation}
      W_{ZF} := H^\dagger,
    \end{equation}
    where $H^\dagger$ denotes pseudo-inverse of $H$.
  \item The eigenspace-LCMV filters~\cite{Sekihara2008} exploiting
    projection of the signal covariance matrix $R$ onto its principal
    subspace of the forms
    \begin{equation}
      W_{EIG{\mhyphen}LCMV(R)} := W_{LCMV(R)} P_{R_{sig}},
    \end{equation}
    and
    \begin{equation}
      W_{EIG{\mhyphen}LCMV(N)} := W_{LCMV(N)} P_{R_{sig}},
    \end{equation}
    where $P_{R_{sig}}$ is the orthogonal projection matrix onto subspace
    spanned by eigenvectors corresponding to $\lambda_1 \geq \dots \geq
    \lambda_{sig}$ --- the $sig$ largest eigenvalues of $R$, where $sig$ is
    the dimension of signal subspace.
  \item The MV-PURE filters, defined as \cite{Piotrowski2019}
    \begin{equation}
      W_{F{\mhyphen}MV{\mhyphen}PURE}^r := P_{L_r^{(i)}} W_{LCMV(R/N)},
    \end{equation}
    for the interference-free model, and~\cite{Piotrowski2019}
    \begin{equation}
      W_{I{\mhyphen}MV{\mhyphen}PURE}^r := P_{K_r^{(i)}} W_{NL},
    \end{equation}
    for the model in presence of interference. In the above expressions,
    $P_{L_r^{(i)}}$, for $i = 1, 2, 3$, are the orthogonal projection
    matrices onto subspaces spanned by eigenvectors corresponding to the
    $r$ smallest eigenvalues of symmetric matrices
    \begin{align*}
      L^{(1)} & := W_{LCMV(R)} R W_{LCMV(R)}^t - 2Q, \\
      L^{(2)} & := W_{LCMV(R)} R W_{LCMV(R)}^t, \\
      L^{(3)} & := W_{LCMV(N)} N W_{LCMV(N)}^t,
    \end{align*}
    respectively; similarly,
    $P_{K_r^{(i)}}$, for $i = 1, 2, 3$, are the orthogonal projection
    matrices onto subspaces spanned by eigenvectors corresponding to the
    $r$ smallest eigenvalues of symmetric matrices
    \begin{align*}
      K^{(1)} & := W_{LCMV(R)} RW_{LCMV(R)}^t - 2Q, \\
      K^{(2)} & := W_{LCMV(R)} R W_{LCMV(R)}^t, \\
      K^{(3)} & := W_{LCMV(N)} N W_{LCMV(N)}^t,
    \end{align*}
    respectively. Here, $Q$ is the covariance matrix of sources of interest $q$.
\end{enumerate}

The file \lstinline!EEGReconstruction.ipynb! is richly commented using
mathematical formulas, thanks to which it will be easy to find a
concrete place of implementation of a specific filter.

\subsection{Configuration}

Table~\ref{tab:setup} contains all configuration parameters. Some of
the more complex parameters have been explained below the table.

\begin{longtable}{|c|p{0.77\textwidth}|}
  \caption{\lstinline!SETUP! configuration.}\label{tab:setup}\\
  \hline
  Parameter & Description \\
  \hline
  \lstinline!rROI! & random (if 1) or predefined (if 0) ROIs \\
  \lstinline!rPNT! & random (if 1) or predefined (if 0) candidate points for source locations \\
  \lstinline!SRCS! & represent \lstinline!SrcActiv!, \lstinline!IntNoise! and \lstinline!BcgNoise!, respectively \\
  \lstinline!DEEP! & deep sources \\
  \lstinline!ERPs! & add ERPs (timelocked activity) \\
  \lstinline!n00! & number of time samples per trial \\
  \lstinline!K00! & number of independent realizations of signal and noise based on generated MVAR model \\
  \lstinline!P00! & order of the MVAR model used to generate time-courses for signal of interest \\
  \lstinline!FRAC! & proportion of ones to zeros in off-diagonal elements of the MVAR coefficients masking array \\
  \lstinline!STAB! & VAR stability limit for MVAR eigenvalues (less than 1.0 results in more stable model producing more stationary signals \cite{2001_Neumaier_Estimationofparametersandeigenmodesofmultivariateautoregressivemodels}) \\
  \lstinline!RNG! & range for pseudo-random sampling of eigenvalues for MVAR coefficients range \\
  \lstinline!ITER! & iterations limit for MVAR pseudo-random sampling and stability verification \\
  \lstinline!PDC_RES! & frequency resolution vector for normalized PDC and DTF estimation \\
  \lstinline!TELL! & provide additional comments during code execution (``tell me more'') \\
  \lstinline!PLOT! & plot figures during the intermediate stages \\
  \lstinline!SCRN! & get screens positions \\
  \lstinline!DISP! & force figures to be displayed on (3dr) screen \\
  \lstinline!SEED! & seed for random number generation \\
  \lstinline!SEEDS! & hard-coded seeds to ensure repeatability of the simulation \\
  \lstinline!RANK_EIG! & rank of EIG-LCMV filter: set to number of active sources \\
  \lstinline!fltREMOVE! & to keep (if 0) or remove (if 1) selected filters \\
  \lstinline!SHOWori! & to show (if 1) or do not show (if 0) Original and Dummy signals on figures \\
  \lstinline!IntLfgRANK! & rank of patch-constrained reduced-rank lead-field \\
  \lstinline!supSwitch! & \lstinline!rec!: run reconstruction of sources activity, \lstinline!loc!: find active sources \\
  \lstinline!thalamus! & type of head model \\
  \lstinline!DEBUG! & if we want to debug \\
  \lstinline!PATH! & path to directory with the code \\
  \lstinline!SRATE! & sampling rate \\
  \lstinline!CUBE! & perturbation of the lead-fields based on the shift of source location within a cube of given edge length (centered at the original lead-fields locations) \\
  \lstinline!CONE! & perturbation of the lead-fields based on the rotation of source orientation (azimuth TH, elevation PHI) \\
  \lstinline!H_Src_pert! & use original (if 0) or perturbed (if 1) lead-field for signal reconstruction \\
  \lstinline!H_Int_pert! & use original (if 0) or perturbed (if 1) lead-field for nulling constrains \\
  \lstinline!SINR! & signal to interference noise power ratio expressed in dB (both measured on electrode level) \\
  \lstinline!SBNR! & signal to biological noise power ratio expressed in dB (both measured on electrode level) \\
  \lstinline!SMNR! & signal to measurment noise power ratio expressed in dB (both measured on electrode level) \\
  \lstinline!WhtNoiseAddFlg! & white noise admixture in biological noise interference noise (FLAG) \\
  \lstinline!WhtNoiseAddSNR! & SNR of \lstinline!BcgNoise! and \lstinline!WhiNo! (dB) \\
  \lstinline!SigPre! & final signal components for pre-interval (use zero or one for signal) \\
  \lstinline!IntPre! & final signal components for pre-interval (use zero or one for interference noise) \\
  \lstinline!BcgPre! & final signal components for pre-interval (use zero or one for background activity) \\
  \lstinline!MesPre! & final signal components for pre-interval (use zero or one for measurement noise) \\
  \lstinline!SigPst! & final signal components for post-interval (use zero or one for signal) \\
  \lstinline!IntPst! & final signal components for post-interval (use zero or one for interference noise) \\
  \lstinline!BcgPst! & final signal components for post-interval (use zero or one for background activity) \\
  \lstinline!MesPst! & final signal components for post-interval (use zero or one for measurement noise) \\
  \lstinline!DATE! & date \\
  \lstinline!NAME! & temporary file name \\
  \lstinline!SINR_RNG! & range of SNR for interference signals\\
  \lstinline!SBNR_RNG! & range of SNR for background signals\\
  \lstinline!SMNR_RNG! & range of SNR for measurment noise\\
  \hline
\end{longtable}

Within \lstinline!SETUP!, perhaps the most important are the
parameters controlling configuration of activity of sources
(\lstinline!SRCS!), configuration of deep sources (\lstinline!DEEP!),
number of samples and realizations of the signal (\lstinline!n00!,
\lstinline!K00!), presence of ERPs (\lstinline!ERPs!), number of
iterations (\lstinline!ITER!), configuration of lead-fields
perturbation (\lstinline!CUBE!, \lstinline!CONE!), signal to noise
ratios, (\lstinline!SINR!, \lstinline!SBNR!, \lstinline!SMNR!), and
presence of signal components (\lstinline!SigPre!, \lstinline!IntPre!,
\lstinline!BcgPre!, \lstinline!MesPre!, \lstinline!SigPst!,
\lstinline!IntPst!, \lstinline!BcgPst!, \lstinline!MesPst!).
  




There are 7 structures containing head conduction model. They are listed in Table~\ref{tab:mats}.

\begin{longtable}{|c|p{0.6\textwidth}|}
  \caption{\lstinline!MATS! structure containing all meshes.}\label{tab:mats}\\
  \hline
  Parameter & Description \\
  \hline
  \lstinline!sel_msh! & structure with head compartments geometry (cortex) \\
  \lstinline!sel_geo_deep_thalami! & structure with mesh containing candidates for location of deep sources (based on \textit{thalami}) \\
  \lstinline!sel_geo_deep_icosahedron642! & structure with mesh containing candidates for location of deep sources (based on \textit{icosahedron642}) \\
  \lstinline!sel_atl! & structure with cortex geometry with (anatomical) ROI parcellation \\
  \lstinline!sel_vol! & structure with volume conduction model (head-model) \\
  \lstinline!sel_ele! & structure with geometry of electrode positions \\
  \lstinline!sel_src! & structure with all cortex lead-fields \\
  \hline
\end{longtable}

{\bf Acknowledgements}. This study was supported by the National Science Center Poland,
grant no. UMO-2016/20/W/NZ4/00354. 

\bibliographystyle{spmpsci}
\bibliography{IEEEabrv,bibliography}

\end{document}